\documentclass[%
 reprint,
]{revtex4-2}

\usepackage{amssymb}
\usepackage{lipsum}
\usepackage{color}
\usepackage{graphicx}
\usepackage{dcolumn}
\usepackage{bm}
\usepackage[utf8]{inputenc}
\usepackage{gensymb}
\usepackage{latexsym}
\usepackage{amsmath}
\usepackage{amsfonts}
\usepackage{nicefrac}
\usepackage{float}
\usepackage{booktabs}
\usepackage{siunitx}
\usepackage{slashed}
\usepackage{hhline}
\usepackage[table]{xcolor}

\usepackage{url}
\usepackage{xspace}
\usepackage{siunitx}
\usepackage{xfrac}
\usepackage{hyperref}
\usepackage[nameinlink]{cleveref}
\usepackage{appendix}
\usepackage{braket}

\usepackage{xifthen}
\usepackage{xcolor}
\hypersetup{colorlinks,	linkcolor={green!55!black},	citecolor={green!60!black},	urlcolor={green!55!black}}

\usepackage{hyperref}
\hypersetup{
    colorlinks,
    citecolor=blue,
    filecolor=blue,
    linkcolor=blue,
    urlcolor=blue
}

\DeclareFontFamily{OT1}{pzc}{}
\DeclareFontShape{OT1}{pzc}{m}{it}%
{<-> s * [1.15] pzcmi7t}{}
\DeclareMathAlphabet{\mathpzc}{OT1}{pzc}{m}{it}

\newcommand{\be}{\begin{equation}}
	\newcommand{\bea}{\begin{eqnarray}}
		\newcommand{\ee}{\end{equation}}
	\newcommand{\eea}{\end{eqnarray}}

\def\1eq#1{Eq.~(\ref{#1})}

\def\2eqs#1#2{Eqs.~(\ref{#1}) and (\ref{#2})}

\graphicspath{{./figures/}{./}}

\begin{document}


\title{Timelike  form factor  for the anomalous process $\gamma^\ast \pi \rightarrow \pi \pi$}

\author{Angel S. Miramontes$^{1}$}
\email{angel.s.miramontes@uv.es}
\author{Gernot Eichmann$^2$}
\email[]{gernot.eichmann@uni-graz.at}
\author{Reinhard Alkofer$^{2}$}
\email[]{reinhard.alkofer@uni-graz.at}

\affiliation{$^1$Department of Theoretical Physics and IFIC, 
University of Valencia and CSIC, E-46100 Valencia, Spain}
\affiliation{$^2$Institute of Physics, University of Graz, NAWI Graz, Universitätsplatz 5, 8010 Graz, Austria}

\begin{abstract}
The form factor $F_{3\pi}(s,t,u)$ for the anomalous process $\gamma^\ast \pi \rightarrow \pi \pi$  
is calculated in the isospin limit for several values of the light current-quark mass (i.e., the pion mass) 
using Dyson–Schwinger and Bethe–Salpeter equations. Beyond a quark interaction 
kernel representing gluon-mediated interactions, leading beyond-rainbow-ladder effects at low 
energies are incorporated by back-coupling pions as explicit degrees of freedom. Building upon an earlier 
calculation of the quark-photon vertex that captures the branch cut associated with  the 
two-pion threshold and the $\rho$-meson resonance, the form factor $F_{3\pi}(s,t,u)$ is determined
for timelike Mandelstam $s$. In particular, predictions are made for the kinematics relevant to the 
Primakoff reaction studied with COMPASS/AMBER at CERN. 
\end{abstract}

\maketitle




\section{Introduction}

Decays and reactions involving an odd number of pions are directly related to the chiral anomaly of QCD.
Among these, the decay of the neutral pion, proceeding almost exclusively via $\pi^0\to \gamma \gamma$, 
has played a historically significant role (cf.~Sec.~11-5 of \cite{Itzykson:1980rh}). 
In the chiral limit, the corresponding decay amplitude is fixed by the electromagnetic and anomalous chiral Ward identities. 
Apart from a numerical factor $1/4\pi^2$, it depends only on the elementary electric charge $e$ and the pion decay constant $f_\pi$,  
and its value has been confirmed experimentally to sub-percent precision~\cite{PrimEx-II:2020jwd}.

The related process $\gamma^\ast \pi \rightarrow \pi \pi$ is similarly constrained  at the soft point in the chiral limit,
but its experimental verification has so far lacked precision. This  is expected to change with the high-precision data
from the ongoing COMPASS analysis~\cite{Friedrich:2023kgv,Ecker:2023qae,JFprivate}, where the kinematics of the Primakoff reaction
 allows access to the amplitude
at timelike Mandelstam  $s$ above the two-pion threshold.
Here, the dependence on the  current-quark mass (and thus  the pion mass)  
is two-fold. In addition to the direct effect on the amplitude, 
one must also consider that the measurement is necessarily
performed away from the chiral limit and the soft point. 
In light of these deviations, 
the relatively large value reported by the Serpukhov experiment  in the 1980s~\cite{Antipov:1986tp} may be less 
surprising. 

The low-energy Primakoff reaction $\gamma^\ast \pi \rightarrow \pi \pi$ provides an excellent testing ground for chiral perturbation theory; 
see~\cite{Moinester:2024lwl} and references therein for a concise summary. 
Recently, a data-driven approach combining dispersion relations  with input from lattice QCD \cite{Niehus:2021iin} has yielded a prediction  
for the cross section of the COMPASS reaction. Although both approaches still involve considerable
uncertainties, it appears that the extracted value of the  $\gamma^\ast \pi \rightarrow \pi \pi$ amplitude 
will exceed the chiral-limit prediction at the soft point. However, a recent extraction of the
anomalous form factor from the process $e^+e^-\to 3\pi$  \cite{Hoferichter:2025lcz} finds 
a somewhat lower value.
Moreover, 
the process $\gamma^\ast \pi \rightarrow \pi \pi$  contributes  to the data-driven evaluation of the muon's 
anomalous magnetic moment  $(g-2)_\mu$.
Given the ongoing discrepancy with experimental results~\cite{Aoyama:2020ynm,Colangelo:2022jxc}, 
a deeper understanding of this anomalous process and its momentum dependence is highly relevant.

In this work, we employ functional methods in the form of Dyson–Schwinger equations (DSEs) and Bethe–Salpeter equations (BSEs) 
to study the $\gamma^\ast \pi \rightarrow \pi \pi$  form factor in the timelike region. 
This approach offers a continuum formulation of QCD that incorporates dynamical chiral symmetry breaking 
and provides a consistent treatment of  bound states and resonances \cite{Jain:1993qh,Maris:1997tm,Maris:1997hd,Maris:2000sk,Alkofer:2000wg,Alkofer:2002bp,Fischer:2008wy,Krassnigg:2008bob,Qin:2011dd,Binosi:2014aea,Rojas:2014aka,Binosi:2016rxz,El-Bennich:2016qmb,Eichmann:2016yit,Xu:2022kng,Miramontes:2019mco,Eichmann:2020oqt,Miramontes:2021xgn,Huber:2020ngt,Huber:2021yfy,Eichmann:2023tjk,Gao:2024gdj}. 
Among other applications, it has recently been applied to the timelike pion form 
factor~\cite{Miramontes:2021xgn} and various two-photon transition form factors~\cite{Eichmann:2017wil,Weil:2017knt,Raya:2019dnh,Eichmann:2019tjk,Eichmann:2024glq}, 
illustrating how vector meson dominance emerges from quark and gluon dynamics.  
The $\gamma^\ast \pi \rightarrow \pi \pi$   form factor has also been  studied in the spacelike regime
within this approach~\cite{Alkofer:1995jx,Cotanch:2003xv,Bistrovic:1999dy,Xing:2024bpj}.

This paper is organized as follows. In Sec.~\ref{sec:amp}, we summarize the kinematics of the
$\gamma^\ast \pi \rightarrow \pi \pi$ amplitude.
In Sec.~\ref{sec:formalism}, we give an overview of the formalism used in the present study.
Our results are presented in Sec.~\ref{sec:numerics}, and we conclude in Sec.~\ref{sec:conclusions}.

\section{The process $\gamma^\ast \pi \rightarrow \pi \pi$} \label{sec:amp}

The  amplitude for the process $\gamma^\ast(Q)\, \pi^\pm(P_2) \rightarrow \pi^0(P_3)\, \pi^\pm(P_4)$ 
can be expressed as a matrix element of the electromagnetic current  
\begin{equation}
J^\mu(x) = e \left( \tfrac{2}{3}\, \bar{u} \gamma^\mu u - \tfrac{1}{3} \,\bar{d} \gamma^\mu d \right),
\end{equation}
which encodes the coupling of the photon to the up and down quarks weighted by their electric charges.
The amplitude $A_{3\pi}^\mu$ involving one incoming and two outgoing pions is then the matrix element of this current between the corresponding hadronic states:  
\begin{equation} \label{eq:A_amplitude}
\begin{split}
A_{3\pi}^\mu (P_2, P_3, P_4) &= \bra{\pi(P_3)\, \pi(P_4)} J^\mu(0) \ket{\pi(P_2)}  \\
&= \varepsilon^{\mu\nu\rho\sigma} P_2^\nu P_3^\rho P_4^\sigma F_{3\pi}(s, t, u) \,.
\end{split}
\end{equation}
The Levi-Civita symbol $\varepsilon^{\mu\nu\rho\sigma}$  
contracts the three independent pion momenta $P_2$, $P_3$ and $P_4$ which are onshell, $P_i^2 = -m_\pi^2$. 
Therefore, the amplitude involves only one
scalar function $F_{3\pi}(s, t, u)$, which is the form factor describing the dynamics of the transition.
It depends on the three Mandelstam variables $s$, $t$, and $u$, which in a Euclidean metric are defined by  $s = -(Q+P_2)^2$, 
$t = -(Q-P_3)^2$ and $u = -(Q-P_4)^2$.
Assuming isospin symmetry, the squared  (off-shell) momentum of the incoming photon is related to the sum of the Mandelstam variables
via $s + t + u = 3m_\pi^2 - Q^2$.

 \begin{figure}[t]
\centerline{%
\includegraphics[width=0.88\columnwidth]{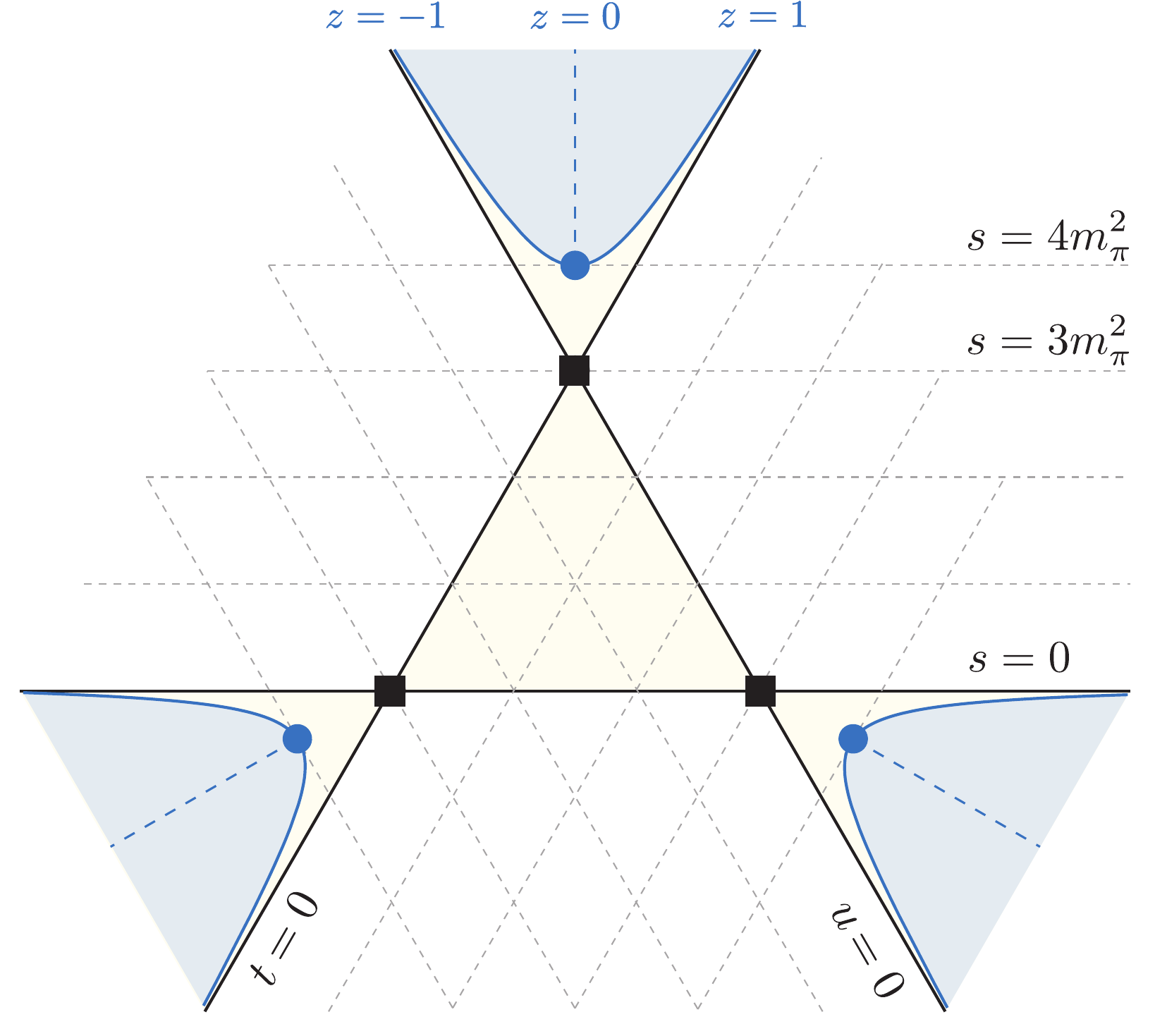}}
\caption{Mandelstam plane for the process $\gamma \pi \rightarrow \pi \pi$. 
 The colored (blue) regions are the physical regions starting at the respective thresholds (blue dots).
 The soft points are shown by the black squares.}
\label{fig:mandelstam}     
\end{figure}

In the following we focus on the soft-photon limit, where $Q^2=0$ and $s + t + u = 3m_\pi^2$.
The corresponding Mandelstam plane in the variables $s$ and $t-u$ is shown in Fig.~\ref{fig:mandelstam}.
The physical $s$-, $t$- and $u$-channel regions are defined by the condition $stu > m_\pi^6$ 
and indicated by the colored areas.
For example, the $s$-channel region 
can be parametrized by the Mandelstam variable $s$, which 
starts at the two-pion threshold $s=4m_\pi^2$,
and the scattering angle $\Theta$ in the center-of-mass (CM) frame,
\begin{equation}\label{scattering-angle}
   z = \cos\Theta =  \frac{t-u}{s-m_\pi^2} \sqrt{\frac{s}{s-4m_\pi^2}} \in [-1,1]\,.
\end{equation}
The boundaries of this region correspond to $z = \pm 1$,
and the  central value $z=0$ is the dashed line in Fig.~\ref{fig:mandelstam}.
In the following we denote the soft points by the kinematic limits where any pair of Mandelstam variables vanishes, e.g. $t=u=0$ and $s=3m_\pi^2$.

The form factor $F_{3\pi}(s, t, u)$ is strongly constrained by
 permutation symmetries. It is invariant under the exchange of any Mandelstam variables, which entails that the Mandelstam plane
is symmetric under a rotation by $120^\circ$ and that the form factor
must be an even function of $z$. Its momentum dependence is governed by  physical poles and cuts.
From Fig.~\ref{fig:mandelstam} it is clear that
the nearest vector-meson poles must also form triangles,   
whose real parts are given by $s=m_\rho^2$, $t=m_\rho^2$ or $u=m_\rho^2$.

The form factor $F_{3\pi}(s,t,u)$ encapsulates the effects of the chiral anomaly and the strong interaction dynamics governing the decay process, making it a fundamental quantity in the study of anomalous QCD processes.
The low-energy behavior of the $\gamma^\ast \pi \rightarrow \pi \pi$ process is governed by the Wess–Zumino–Witten (WZW) anomaly \cite{Adler:1971nq,Aviv:1971hq}. At leading order in the chiral expansion, the amplitude is completely fixed by symmetry considerations and depends only on the pion decay constant and the electric charge. 
Using $e=\sqrt{4\pi \alpha_\text{em}}  \approx 0.3028$ and the physical value of the pion decay constant, $f_{\pi} = 92.28(10)~\text{MeV}$, 
the corresponding value of the anomalous form factor in the chiral limit and at the
soft point ($s=t=u=0$) is 
\begin{equation}
    F_{3 \pi}^{\text{anomaly}}= \frac{e}{4 \pi^2 f_{\pi}^3}  = 9.76(3)~\text{GeV}^{-3} \,.
    \label{eq:anomaly}
\end{equation}
We will return to this expression in Sec.~\ref{sec:numerics}.

\section{Dyson–Schwinger/Bethe–Salpeter formalism }
\label{sec:formalism}

In this section we present  key aspects of the  DSE/BSE framework used in our study;
details can be found in the reviews~\cite{Alkofer:2000wg,Eichmann:2016yit,
Bashir:2012fs,Sanchis-Alepuz:2017jjd}.
We employ a Euclidean metric and, if necessary, perform analytic continuations to timelike momenta via contour deformations.

\begin{figure}[t]
\centerline{%
\includegraphics[width=1\columnwidth]{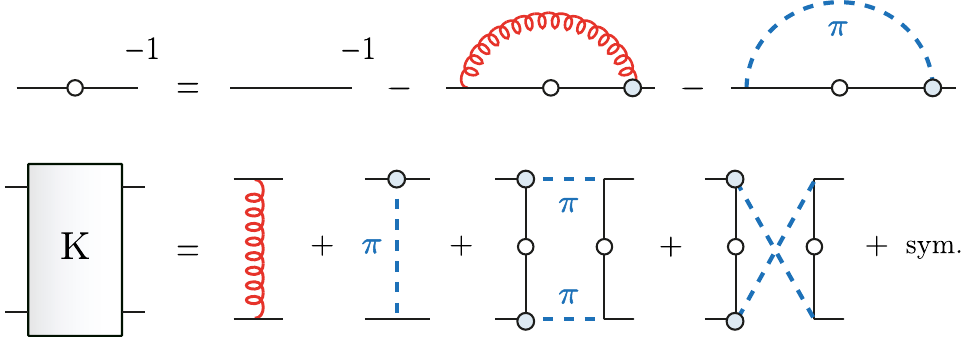}}
\caption{Quark propagator (top) and interaction kernel (bottom) within the beyond-rainbow-ladder truncation employed here,
including explicit pion exchange contributions.}
\label{fig:BRL_truncation}     
\end{figure}

The dressed quark propagator $S(p)$ incorporates dynamical chiral symmetry breaking, a key feature of QCD that significantly influences the structure of hadrons.  For a given flavour $f$ it can be obtained by solving the corresponding DSE
\begin{equation}\label{quark-dse}
\begin{split}
S^{-1}(p) &= Z_2\,(i\slashed{p} + Z_m m_f) \\
&- Z_{1f} \,g^2 \,C_F\int_q i\gamma^\mu  \, S(q) \, \Gamma^\nu_\text{qg}(q,p)\,D^{\mu \nu}(k) \,,
\end{split}
\end{equation}
where $\int_q = \int d^4 q/(2\pi)^4$, $m_f$ is the renormalised current-quark mass that enters in the QCD Lagrangian, 
$C_F=4/3$ is the Casimir in the fundamental representation of $SU(N_c=3)$,
$k=q-p$ is the gluon momentum, 
$D^{\mu \nu}$  the dressed gluon propagator, $\Gamma^\nu_\text{qg}$ the dressed quark-gluon vertex,
and $Z_2$, $Z_m$ and $Z_{1f}$ are renormalisation constants.
The  DSE is shown diagrammatically  in the top panel of 
Fig.\ \ref{fig:BRL_truncation}; the last diagram with the pion exchange will be
discussed below.
In the following we  restrict ourselves to the  light $u$, $d$ flavours in the isospin limit.

Mesons, on the other hand, are described as bound states or resonances of a quark and an antiquark.
For example, the Bethe–Salpeter amplitude $\Gamma_\pi$ of the pion is obtained as a solution of the homogeneous BSE
\begin{equation}  \label{eq:homogeneousBSE} 
\begin{split}
[\Gamma_\pi(p,P)]_{\alpha\beta} &= C_F \int_q \, [ \mathbf{K}(p,q,P)]_{\alpha\gamma;\delta\beta}   \\
& \quad \times [S(q_+)\, \Gamma_\pi(q,P)\,S(q_-)]_{\gamma\delta}\,, 
\end{split}
\end{equation}
where $P$ is the total momentum, $p$  the relative momentum between the quark and antiquark, $q$ the relative momentum in the loop, and $q_\pm = q \pm P/2$.
The Bethe-Salpeter kernel $\mathbf{K}$ encapsulates the dynamics of the quark-antiquark interaction,
and the bottom panel of Fig.~\ref{fig:BRL_truncation} is the version used herein.

The internal structure of a meson can be explored through the interaction of a photon with the quarks it contains. This  is described by the  dressed quark-photon vertex $\Gamma^{\mu}$, which plays a fundamental role in processes involving electromagnetic transitions and hadronic structure \cite{Frank:1994mf,Maris:1999bh,Eichmann:2014qva, Miramontes:2019mco,Leutnant:2018dry,Tang:2019zbk}. The quark-photon vertex is determined by solving the inhomogeneous BSE, 
\begin{equation}  \label{eq:inhomBSE_vector}
\begin{split}
[\Gamma^{\mu}(p,Q)]_{\alpha\beta} &= Z_2\,i\gamma^\mu_{\alpha\beta}\,\mathsf{Q} + C_F \int_q \, [ \mathbf{K}(p,q,P)]_{\alpha\gamma;\delta\beta}  \\
& \quad \times [S(q_+)\, \Gamma^\mu(q,P)\,S(q_-)]_{\gamma\delta}\,, 
\end{split}
\end{equation}
with the same ingredients as in Eq.~\eqref{eq:homogeneousBSE} 
except $Q$ is now the photon momentum and $q_\pm = q \pm Q/2$.
The additional inhomogeneous term contains the quark electric charge matrix $\mathsf{Q} = \text{diag}(\frac{2}{3},-\frac{1}{3})$
for up and down quarks,
which determines the isospin structure of the vertex.

The quark-photon vertex can be expressed in a Lorentz-Dirac tensor basis, which 
leads to two distinct components: $\Gamma^\mu(p,Q) = \Gamma^\mu_T(p,Q) + \Gamma^\mu_\text{BC}(p,Q)$, 
where $\Gamma^\mu_T$ depends on eight tensors that are transverse to the photon momentum $Q^\mu$ 
and the Ball-Chiu vertex $\Gamma^\mu_\text{BC}(p,Q)$ consists of four non-transverse components, see, e.g.,~\cite{Ball:1980ay,Alkofer:2000wg}.
The latter are fully determined by the quark propagator through  the electromagnetic
Ward-Takahashi identity, which is crucial for preserving electromagnetic gauge invariance, 
while the transverse terms ensure a consistent description of bound-state dynamics.  
In this work, we incorporate all twelve components in our calculations 
to provide a complete and self-consistent treatment of the quark-photon interaction.
In particular, the transverse part includes the mixing of off-shell photons with the $\rho$ resonance
and dynamically generates vector-meson poles in the quark-photon vertex,
while simultaneously developing a two-pion cut that moves the  
$\rho$-resonance pole on the second Riemann sheet~\cite{Miramontes:2019mco}.

Without the pion terms in the DSE and in the interaction kernel, 
our truncation  shown in Fig. \ref{fig:BRL_truncation} would correspond to the widely used 
rainbow ladder (RL) truncation. In the quark DSE this entails
\begin{equation}
Z_{1f}\,g^2\,\Gamma^\nu_\text{qg}(q,p)\,D^{\mu \nu}(k) \; \to \; Z_2^2\,\frac{4\pi\alpha(k^2)}{k^2}\,T^{\mu\nu}_k\,i\gamma^\nu\,,
\end{equation}
where $T^{\mu \nu}_k = \delta^{\mu\nu} - k^\mu k^\nu/k^2$ is a transverse projector
and $\alpha(k^2)$ an effective interaction that incorporates dressing effects of the gluon propagator and the quark-gluon vertex. 
The kernel $\mathbf{K}$ in the BSEs (\ref{eq:homogeneousBSE}--\ref{eq:inhomBSE_vector}) is then uniquely related to the
quark-self energy by vector and axial-vector Ward-Takahashi identities and given by
\begin{equation}
[ \mathbf{K}(p,q,P)]_{\alpha\gamma;\delta\beta} \; \to \; Z_2^2\,\frac{4\pi\alpha(k^2)}{k^2}\, i\gamma^\mu_{\alpha\gamma}\,T^{\mu\nu}_k\, i\gamma^\nu_{\delta\beta}\,.
\end{equation}
This construction satisfies chiral constraints like the Gell-Mann-Oakes-Renner relation and ensures the
\mbox{(pseudo-)} Goldstone boson nature of the pion. Specifically, in the Maris–Tandy model~\cite{Maris:2000sk} the effective interaction takes the form 
$\alpha(k^2) = \pi \eta^7 x^2\,e^{-\eta^2 x} + \alpha_\text{UV}(k^2)$ 
with $x=k^2/\Lambda^2$, where $\Lambda$ is a scale and $\eta$  a width parameter.
The term $\alpha_\text{UV}(k^2)$ ensures the correct perturbative running but is otherwise not important for low-energy properties.

Beyond the rainbow-ladder truncation, additional diagrams contribute to the interaction kernel  both in the DSEs and BSEs.
Here, certain important classes of quark and gluon diagrams~\cite{Fischer:2007ze,Fischer:2008sp,Fischer:2008wy} can be rearranged to yield the expression
in Fig.~\ref{fig:BRL_truncation}, which involves effective pion-exchange interactions. 
{These contributions correspond to the approximate next-to-leading order corrections in a 1/$N_c$ expansion, where the pion exchange emerges as the leading hadronic effect beyond one-gluon exchange. While Eq.~\eqref{quark-dse}  contains {to leading order} the gluon-mediated interaction {only}, the additional pion exchange in Fig.~\ref{fig:BRL_truncation} arises from the resummation of non-Abelian diagrams \cite{Fischer:2007ze}. }

{In particular, the pion-exchange interactions} incorporate the  decay channel of a vector meson into two pions~\cite{Williams:2018adr,Miramontes:2019mco}. 
This mechanism, which includes the non-trivial 
analytic structure due to intermediate hadron resonances, is essential for an accurate description of the electromagnetic pion form factor in the timelike momentum region up to and beyond the $\rho$ resonance, as it accounts for the resonant structure of the process and ensures consistency with experimental observations~\cite{Miramontes:2021xgn,Alkofer:2022hln}.  The technical details of the truncation are discussed in depth elsewhere \cite{Williams:2018adr,Miramontes:2019mco}.
Note that the coupling of the pion to the quark in Fig. \ref{fig:BRL_truncation} 
is calculated self-consistently from the pion BSE in Eq.~\eqref{eq:homogeneousBSE},
which thus becomes a non-linear integral equation.

\begin{figure}[t]
\centerline{%
\includegraphics[width=1\columnwidth]{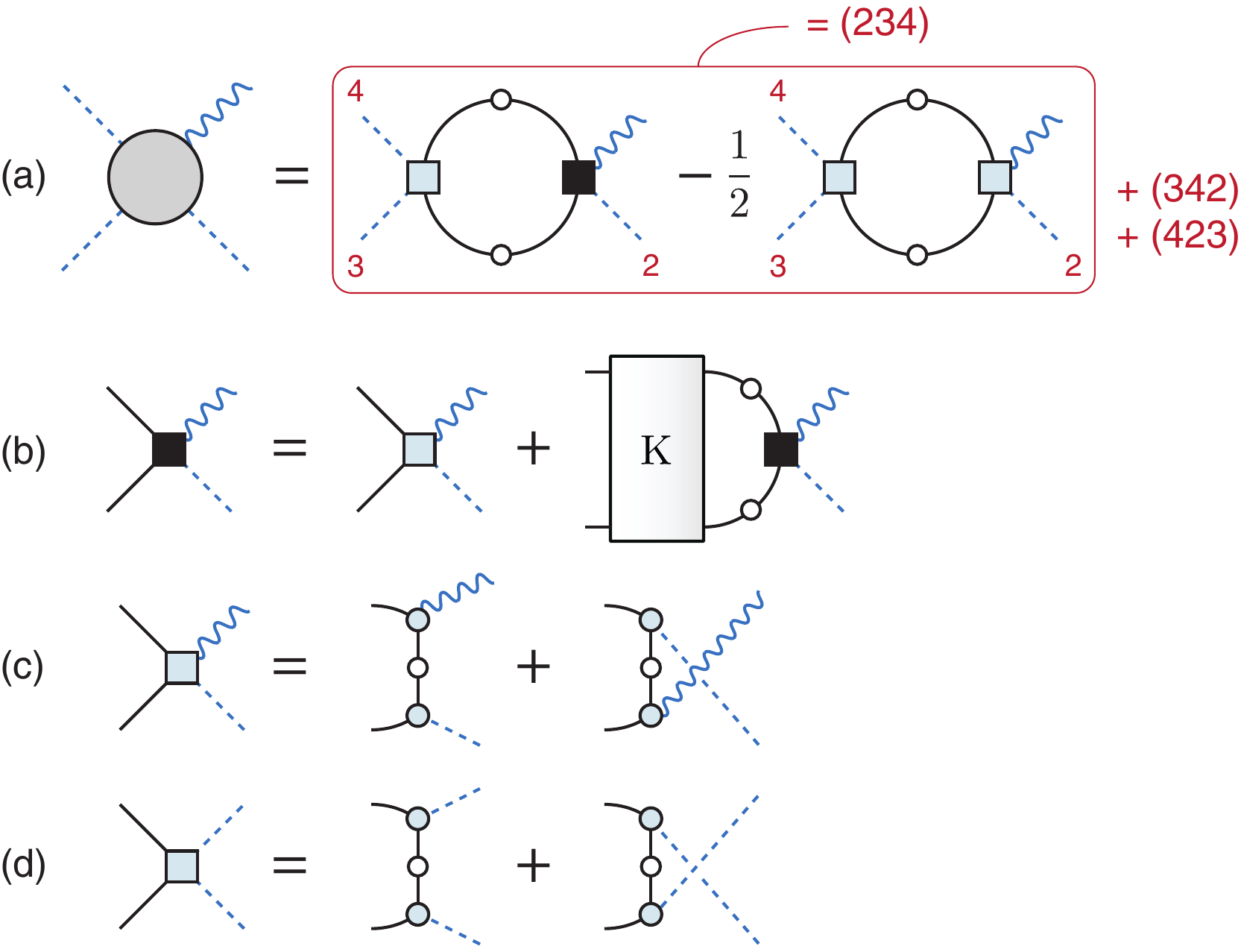}}
\caption{(a) $\gamma^\ast \pi \rightarrow \pi \pi$ matrix element;
         (b) inhomogeneous Bethe-Salpeter equation for the $\gamma^\ast \pi \to q\bar{q}$ vertex,
         (c) $\gamma^\ast \pi \to q\bar{q}$ Born terms, and
         (d) $\pi \pi \to q\bar{q}$ Born terms. }
\label{fig:4pt}     
\end{figure}

Let us now turn to the calculation of the process $\gamma^\ast \pi \rightarrow \pi \pi$.
Like other four-point functions~\cite{Cotanch:2002vj,Goecke:2012qm,Eichmann:2011ec,Eichmann:2012mp,Eichmann:2014ooa}, this matrix element can be derived by repeated functional derivatives 
to yield the system of equations shown in Fig.~\ref{fig:4pt}.
The first line (a) is the result for the $\gamma^\ast \pi \rightarrow \pi \pi$ amplitude.
It depends on the full $\gamma^\ast \pi \to q\bar{q}$ vertex $G^{\mu}(p,Q,P_2)$ denoted by the black square.
This quantity satisfies the inhomogeneous BSE in line (b),
\begin{align}  \label{eq:fullvertex}
& [G^{\mu}(p,Q,P_2)]_{\alpha\beta} = [G_0^{\mu}(p,Q,P_2)]_{\alpha\beta}   \\
& + \int_q \, [ \mathbf{K}(p,q,P)]_{\alpha\gamma;\delta\beta}  [S(q_+)\, G^{\mu}(q,Q,P_2)\,S(q_-)]_{\gamma\delta}\,,  \nonumber
\end{align}
where $Q$ and $P_2$ are the incoming photon and pion momenta, $P = Q+P_2$, $q_\pm = q \pm P/2$,
and we suppressed the color and flavor factors in the notation.
The BSE is structurally analogous to Eq.~\eqref{eq:inhomBSE_vector} 
except that the inhomogeneous term is given by the $\gamma^\ast \pi \to q\bar{q}$ Born diagrams $G_0^{\mu}(p,Q,P_2)$ shown in line (c).
Here, the photon and  pion couple to the quark via the quark-photon vertex and the onshell pion amplitude, respectively,
\begin{equation}\label{born}
\begin{split}
   G_0^\mu(p,Q,P_2) &= \Gamma_\pi(l_+,P_2)\,S(p+K)\,\Gamma^\mu(r_-,Q) \\
                &+ \Gamma^\mu(r_+,Q)\,S(p-K)\,\Gamma_\pi(l_-,P_2)\,,
\end{split}
\end{equation}
with $K = (Q-P_2)/2$, $l_\pm = p \pm Q/2$ and $r_\pm = p \pm P_2/2$.
The quantities in Eq.~\eqref{born} are determined beforehand by Eqs.~(\ref{quark-dse}--\ref{eq:inhomBSE_vector}).
The $\pi \pi \to q\bar{q}$ Born vertex in line (d) is analogous and follows from Eq.~\eqref{born}
by replacing the quark-photon vertex with the pion amplitude.

Through Eq.~\eqref{eq:fullvertex}, the $\gamma^\ast \pi \to q\bar{q}$ vertex
contains an infinite set of gluon ladder diagrams, which ensures a consistent treatment of nonperturbative effects.
Moreover, this automatically
generates all meson poles in the $\gamma^\ast \pi \to q\bar{q}$ channel.
After putting the vertex back in line~(a), 
the intermediate $s$-, $t$- and $u$-channel meson poles in the $\gamma^\ast \pi \rightarrow \pi \pi$ amplitude are therefore automatic.
Note that these poles would be lost when replacing $G^\mu \to G_0^\mu$, i.e.,
when replacing the full vertex with the Born terms, 
in which case the amplitude turns into a  sum of  quark box diagrams.
While this  captures the leading-order contributions to the process, 
the corresponding anomalous form factor fails to reproduce the low-energy theorem  as demonstrated in \cite{Cotanch:2003xv}.
Thus, the gluon ladders play a crucial role in preserving electromagnetic gauge invariance and chiral symmetry.  
We also note that the same BSE (b, c) is involved in the calculation of pion electroproduction on the nucleon~\cite{AnDiPrep}, 
and the similar equation with two photons enters in nucleon Compton scattering and the hadronic light-by-light scattering amplitude~\cite{Eichmann:2012mp,Eichmann:2014ooa};
the latter has the same form as in line (a) except with four photons.

The resulting expression inside the rectangular box in line (a) has the  structure of Eq.~\eqref{eq:A_amplitude} 
but with $F_{3\pi}(s,t,u)$  replaced by a function $f(s,t,u)$.
The full result is then obtained by summing up the permutations of the pions:
\begin{equation}
F_{3\pi}(s, t, u) = f(s, t, u) + f(t, u, s) + f(u, s, t).
\end{equation}  

While we employ the full BSE kernel in Fig. \ref{fig:BRL_truncation} in the calculation of the quark propagator, 
the meson amplitudes and the quark-photon vertex, the computational demand for including the two-pion exchange terms 
in the calculation of the $\gamma \pi \to q\bar{q}$ vertex is extremely large and therefore left for future work.
Here we only implement the additional single-pion exchange diagrams.
As a consequence, only the vector-meson poles
at timelike values of the photon virtuality $Q^2$ are resonance poles on the second Riemann sheet,
while the $s$-, $t$- and $u$-channel meson poles
 lie on the real axis.
In Sec.~\ref{sec:numerics} below we will use the fact that we can switch off the resonance mechanism 
in the kernel to estimate the effect of the $\rho$-meson width.
Strictly speaking, the equations in Fig.~\ref{fig:4pt} are also only valid in rainbow-ladder,
because in general the photon can also couple to the BSE kernel~\cite{Eichmann:2011ec}.
Although this introduces a slight inconsistency, we will see in Sec.~\ref{sec:numerics} that the resulting discrepancies are very small.

\begin{table}[t]
    \centering
    \renewcommand{\arraystretch}{1.2} 
    \setlength\tabcolsep{2.5mm}
    \begin{tabular}{c c c c c}  
        \toprule
        & $m_{\pi}$[MeV] & $f_{\pi}$  [MeV]  & $m_{\rho}$  [MeV] & $f_{\rho}$  [MeV]  \\
        \midrule
        BRL  & 139(2) & 92(1) & 748(5) & 159(2) \\
        Exp. & 139    & 92.3(1) & 775.3(3) & 156(1) \\
        \bottomrule  
    \end{tabular}
    \caption{Masses and decay constants computed in the beyond-rainbow-ladder (BRL) truncation
(first line). The  errors are obtained by varying the interaction strength parameter 
$\eta = 1.65 \pm 0.05$. In the second line the corresponding  experimental values are given \cite{ParticleDataGroup:2024cfk}.}
    \label{tab:masses_and_decay_constants}
\end{table}

\begin{figure*}[t]
\centerline{%
\includegraphics[width=1\textwidth]{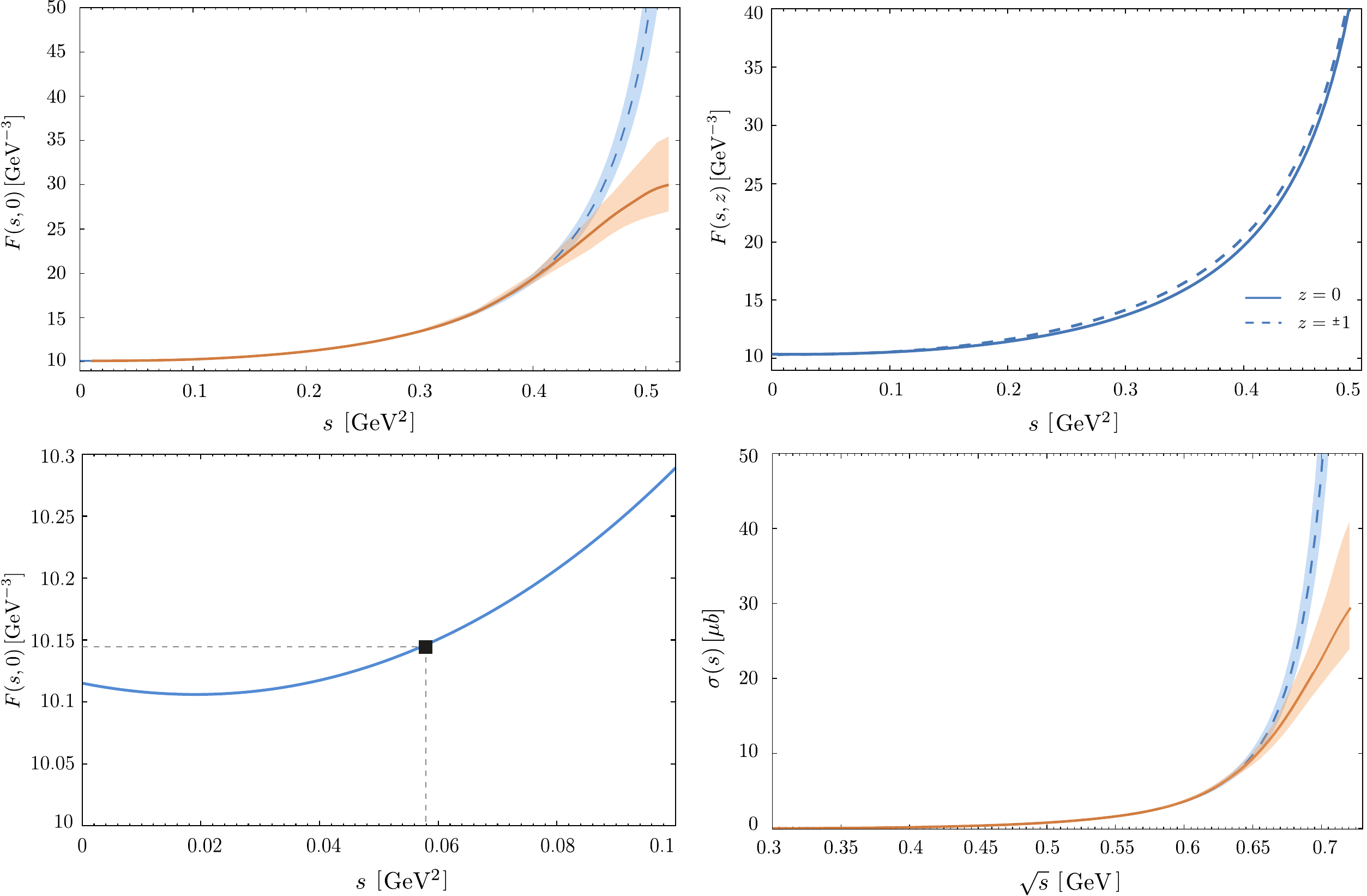}}
\caption{Anomalous form factor $F(s,z)$ for the COMPASS kinematics.
 Top left: Numerical result for the form factor of the Primakoff process 
$\gamma \pi \rightarrow \pi \pi$ (dashed blue line) and predicition 
 after subtracting the overestimate due to $\rho$-meson pole (solid orange line).
The errors correspond to the variation of the interaction parameter $\eta$. 
Top right: Numerical result for three different values of the scattering angle
$\Theta$, $z=\cos \Theta = 0$ and $z = \pm 1$. 
Bottom left: Form factor close to the soft point $s=3m_\pi^2$, which is marked by the square. 
Bottom right: Calculated (dashed blue line) and predicted (solid orange line) 
cross section. The error bands correspond to the variation of the interaction parameter $\eta$. }
\label{fig:FF_compass}     
\end{figure*}

\section{Results}
\label{sec:numerics}

To calculate the form factor in the timelike region, we adopt the kinematics of the 
 COMPASS experiment runs performed in 2009 and 2012 \cite{JFprivate}. 
The incoming 190 GeV $\pi^-$ mesons scatter off a Ni target to produce the two outgoing pions in a Primakoff reaction. 
This implies a small photon virtuality  $Q<$ 36 MeV, with the Primakoff peak located at 12 \ldots 15 MeV but extending 
down to 1 MeV. For negligible values of $Q^2$ the constraint $s+t+u\approx 3m_\pi^2$ applies.
The physical region is then the $s$-channel region in Fig.~\ref{fig:mandelstam}, whose
 kinematics is defined by the Mandelstam variable $s$ and $z=\cos \Theta$ from  Eq.~\eqref{scattering-angle},
from where the Mandelstam variables $t$ and $u$ can  be reconstructed.
Data have been taken for values of $s$ from the threshold $s=(2m_\pi)^2$ up to $(12m_\pi)^2 \approx 2.8 \,\mathrm{GeV}^2$. 
In the following  we abbreviate
the form factor by $F_{3\pi}(s,t,u) = F(s,z)$. 

In our calculations, the light $u/d$ current-quark mass is an input and fixed to $m_{u,d} = 3.7$ MeV at the renormalisation scale $\mu = 19$ GeV.
We set the parameters of the Maris-Tandy interaction to $\Lambda = 0.78$ GeV and $\eta = 1.65 \pm 0.05$ to reproduce the pion mass and decay constant (see Table \ref{tab:masses_and_decay_constants}).  Note also that the 
$\rho$-meson mass and decay constant are fairly well reproduced.
In order to solve the corresponding integral equations needed for the quark DSE and the various BSEs we employ  standard algorithms, 
see~\cite{Eichmann:2016yit, Sanchis-Alepuz:2017jjd,Eichmann:2012mp}
for detailed descriptions of the numerical techniques. 

As explained at the end of Sec.\ \ref{sec:formalism}, when neglecting the two-pion exchange diagrams in the resummed 
ladders the $\rho$-meson decay channel into two pions is closed. 
Thus the $\rho$ meson  in the $s$-channel will not  show up  as a finite-width resonance but  produce a pole in 
$F(s,z)$ at $s=m_\rho^2$ which, based on our calculated values, is located at 
$s\approx (0.75 \,\mathrm{GeV})^2 = 0.56 \,\mathrm{GeV}^2$.
Until we obtain results from a fully self-consistent treatment of the two-pion exchange terms (which will take time 
due to the enormous computational demand) we adopt the following procedure: 
We compare the self-consistently calculated timelike electromagnetic pion form factor  
with a corresponding result that neglects the two-pion exchange terms in the pion form factor's $s$-channel.
From the ratio of these two results we extract correction factors as functions of $s$ and
apply them to the calculated value of $F(s,z)$.
Naturally, the reliability of this estimate is limited to values of the Mandelstam variable $s$ below $m_\rho^2$.

Our main numerical results are summarised in Fig.~\ref{fig:FF_compass}.
The top left panel shows the calculated form factor $F(s,0)$ 
(dashed blue line), along with an error estimate obtained by varying the 
interaction parameter $\eta$ that controls the shape of the effective quark-gluon coupling. 
As expected and discussed above, the resulting form factor diverges at $s=m_\rho^2$.
Applying the correction inferred from the pion form factor yields the solid orange line
and the associated uncertainty band, 
which illustrates the sensitivity of our results to the model parameters and 
largely captures the uncertainty from the absence of the $\rho$ decay in the $s$ channel.

This  reflects only one source of uncertainty;
further systematic errors are difficult to quantify but are expected to be of a similar magnitude.
In particular, for $s < 0.45\ \mathrm{GeV}^2$ the uncertainties are reasonably small, 
which is expected because the anomalous form factor is strongly constrained by symmetries.

In the top right panel of Fig.\ref{fig:FF_compass}, we also show the form factor computed 
for different values of the angular parameter $z$, which reflects its dependence on $t$ and $u$ as defined in Eq.~\eqref{scattering-angle}.
Note that the calculated values for any other value of the scattering angle lie between the full and dashed lines.
In particular, for $s < 0.45\ \mathrm{GeV}^2$, which is the region where our results are expected to be reasonably precise,
the dependence on the scattering angle is comparable to or even smaller than the estimated uncertainties.
The prediction that the form factor is nearly independent of the scattering angle is a surprising outcome of our calculation.
It is exemplary for planar degeneracies seen in other systems that are strongly constrained by permutation symmetries~\cite{Eichmann:2014xya,Eichmann:2014ooa,Eichmann:2015nra,Eichmann:2017wil,Pinto-Gomez:2022brg,Ferreira:2023fva,Aguilar:2023qqd,Aguilar:2024fen}.
It will be interesting to see whether this behaviour can be confirmed experimentally.

In the bottom left panel of Fig.~\ref{fig:FF_compass}, we zoom in on small values of $s$.
As expected from the permutation symmetry in the Mandelstam plane (Fig.~\ref{fig:mandelstam}), 
the form factor $F(s,0)$ decreases for small $s$, reaches a shallow minimum at $s = m_\pi^2 \approx 0.02\ \mathrm{GeV}^2$,
and increases again towards $s=0$.

To allow for a more direct comparison of our theoretical predictions with experimental measurements, 
we compute the total cross section for the $\gamma^\ast \pi \rightarrow \pi \pi$ process
from the values of $F(s,0)$ displayed in the top left panel of Fig.~\ref{fig:FF_compass}. 
With $F(s,z) \approx F(s,0)$, the cross section can be written as~\cite{Niehus:2021iin,Briceno:2016kkp,Hoferichter:2012pm,Hoferichter:2017ftn}
    \begin{equation}
    \sigma(s) = \frac{(s-4 m^2_{\pi})^{3/2}(s-m^2_{\pi})}{768 \pi \sqrt{s}} |F(s,0)|^2 \, .
    \end{equation}
Our prediction as a function of $\sqrt{s}$ is displayed in the lower right panel of Fig.~\ref{fig:FF_compass}. The blue band corresponds to our numerical results, and the orange band represents our prediction including
the uncertainty due to the omission of the decay mechanism in the $s$ channel.

\begin{table}[t]
    \centering
    \renewcommand{\arraystretch}{1.2} 
    \begin{tabular}{c c c c c c}  
        \toprule
        $m_q$ [MeV] & $m_{\pi}$ [MeV] & $f_{\pi}$ [MeV] & $F_{3\pi}^\text{sp}$ [GeV$^{-3}$] & $4 \pi^2 f_{\pi}^3 \,F_{3\pi}^\text{sp}/e$ \\
        \midrule
        0.1 & 28 & 89.8 & 10.65 & 1.0053  \\  
        0.5 & 52 & 90.1 & 10.57 & 1.0083   \\
        1.0 & 74 & 90.4 & 10.52 & 1.0124  \\
        2.0 & 104 & 90.9 & 10.40 & 1.0180  \\                
        3.7 & 139 & 91.9 & 10.14 & 1.0281  \\
        \bottomrule  
    \end{tabular}
    \caption{Pion mass,  pion decay constant and 
anomalous form factor at the soft point, unnormalised and normalised, for different quark masses.}
    \label{tab:anomaly_dependence}
\end{table}

From a theoretical perspective it is, of course, interesting to understand the behaviour of the anomalous 
form factor at the soft point when approaching the chiral limit.
In this context, we note that within the DSE/BSE formalism, a rewriting of the relevant integrals 
at the soft point and in the chiral limit ($s=t=u=0$) allows one
to express the non-vanishing contributions as integrals over total derivatives, see, e.g., \cite{Alkofer:1995jx,Cotanch:2002vj}, 
which can be performed analytically.
In a symmetry-preserving truncation, the result then reproduces the anomaly value.
In particular, the linear dependence on $N_c$ can be traced back to the quark loops in the equations. 

Table~\ref{tab:anomaly_dependence} lists the computed values of the pion masses at different current-quark masses, 
the corresponding decay constants, and the anomalous form factor at the soft point  
within this truncation. To this end, we use the value $F_{3\pi}^\text{sp} = F(3m_\pi^2, 0)$ and send $m_\pi \to 0$
by repeating the calculations for different current-quark masses close to the chiral limit.
To understand the behaviour of $F_{3\pi}^\text{sp}$, it is important to first recognize 
that the anomalous form factor is proportional to $f_\pi^{-3}$ due to the normalization 
of the three pion Bethe–Salpeter amplitudes appearing in the matrix element.
As is evident from Table~\ref{tab:anomaly_dependence}, the pion decay constant in the chiral limit 
is significantly smaller than its physical value. In fact, our estimate for this number
is in close agreement with determinations from chiral perturbation theory, which yields 
$f_\pi = f_0 \left(1 + \mathcal{O}(m_q)\right)$~\cite{Ecker:1994gg}.
This {estimated} 2.4\% decrease in $f_\pi$ from the physical point to the chiral limit 
leads, due to the cubic dependence, to an approximately 7\% increase in the amplitude.
The direct contribution to the matrix element, however, decreases from the physical point towards 
the chiral limit by a few percent, resulting in a net increase of approximately 5\%: 
{The extrapolated chiral limit value of the numerical
result is given by $F_{3\pi}^\text{anomaly} \approx 10.66~\text{GeV}^{-3}$ as compared to 
$F_{3\pi}^\text{soft} \approx 10.14~\text{GeV}^{-3}$ at the physical pion mass.}

\begin{figure}[t]
\centerline{%
\includegraphics[width=0.5\textwidth]{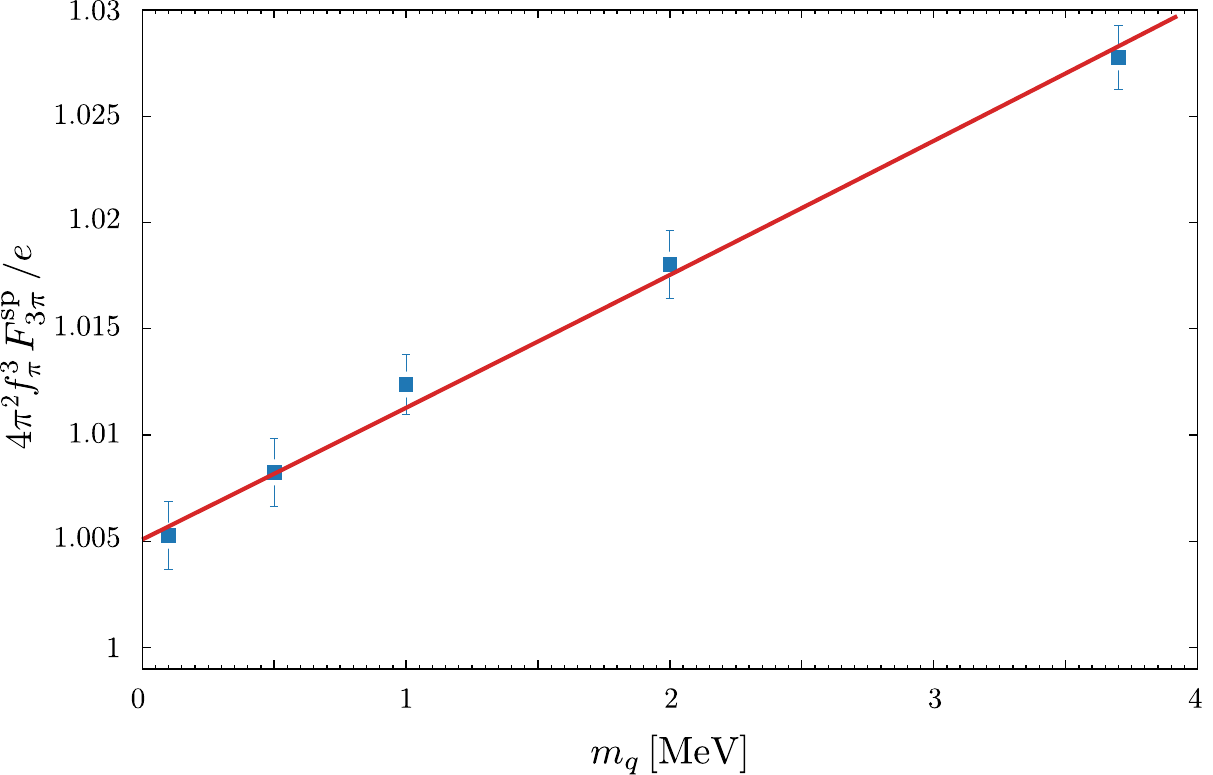}}
\caption{Normalised anomalous form factor at the soft point for different  current-quark masses. The error estimates refer 
to the variation of the interaction strength parameter $\eta$ only.
}
\label{fig:FF_anomaly}     
\end{figure}

In Fig.~\ref{fig:FF_anomaly}, the normalised anomalous form factor at the soft point is displayed for small values of the renormalised 
current-quark mass $m_{u,d}(\mu = 19\,\mathrm{GeV})$. It is straightforward to understand why the error estimates, 
obtained from varying the interaction strength parameter $\eta$, are very small. 
In a symmetry-preserving truncation, the chiral and electromagnetic Ward identities would ensure that the normalised form factor 
assumes the value 1 in the chiral limit, regardless of the details of the interaction. 
The deviation observed here can therefore be attributed to the omission of the two-pion exchange terms in the resummed exchange ladders,
whereas in the quark propagator, the pion amplitude and the quark-photon 
vertex these are included. We therefore conclude that this inconsistency in the calculation
leads to an overestimate of the form factor at the soft point by approximately 0.5\%, which has to be accounted for as an additional systematic error.

{To summarise the effect of the physical current quark mass, 
i.e., the effect of explicit chiral symmetry breaking, we note that 
the total change in our calculation from the chiral limit to the physical point,
corresponds to an increase of the normalised 
 $F_{3\pi}$ at the soft point by 2.8\% $\pm$ 0.5\% $\pm$ 0.2\% $\pm$ 0.2\%.
These uncertainties reflect the above mentioned
errors due to the partial omission of the two-pion exchange kernel,
uncertainties due to the variations in the effective gluon-mediated interaction, and 
other numerical errors.
For comparison, a respective dispersive analysis including chiral logarithms and resonance saturation 
estimates of low-energy constants yields a 6.6(1.0)\% increase from the chiral limit to the physical point 
\cite{Bijnens:1989ff,Hoferichter:2012pm}.
Clearly, these two estimates of the effect of explicit chiral symmetry breaking are at an
approximate 2$\sigma$ tension to each other.}

\section{Conclusions} \label{sec:conclusions}
We  calculated the form factor $F_{3\pi}(s,t,u)$ for the anomalous process
$\gamma^\ast \pi \rightarrow \pi \pi$ in the isospin limit for several values of the light current-quark mass, 
or equivalently the pion mass, by employing Dyson–Schwinger and Bethe–Salpeter equations
in a beyond-rainbow-ladder truncation. We  focused  on timelike values
of the Mandelstam parameter $s$ and determined the values of $t$ and $u$ to adapt
to the kinematics of the Primakoff reaction studied at COMPASS.

Our main findings can be summarised as follows:\\
 (i) The anomalous form factor rises significantly
 between the soft point and even moderate values of $s$.
Correspondingly, the cross section increases significantly for values of $\sqrt{s}$  between the
threshold and the $\rho$ meson mass. \\
(ii)  The anomalous form factor's dependence on the scattering angle is surprisingly small.

As a next step, it will be interesting to include the two-pion exchange terms in a fully self-consistent manner. 
Despite the significant computational cost involved, this effort is worth pursuing -- 
not primarily to reduce the existing uncertainties, but rather to enable calculations 
for $\sqrt{s}$ at and beyond the $\rho$-meson mass. Such results can then be confronted 
with predictions from vector meson models, analogous to the comparison performed 
for the timelike pion form factor in Ref.~\cite{Miramontes:2021xgn}. 
It will be interesting to see whether the anomalous form factor follows similar patterns.
These investigations, along with a study of the anomalous form factor beyond the 
Primakoff kinematics, particularly for the process $e^+e^- \to 3\pi$, will be presented elsewhere.

Last but not least, we would like to emphasize that the present calculation 
constitutes an important first step toward the computation of hadronic $2 \to 2$ scattering amplitudes 
within the Dyson–Schwinger–Bethe–Salpeter framework using a beyond-rainbow-ladder truncation.

\bigskip

\textbf{Acknowledgements}
~\\
 {RA thanks Jan Friedrich and Dominik Ecker from the COMPASS collaboration 
{as well as Martin Hoferichter} for helpful discussions. 

 ASM is supported by the “PROMETEO” programme of the “Generalitat Valenciana” grant CIPROM/2022/66, and the Spanish “Agencia Estatal de Investigación”, MCIN/AEI/10.13039/501100011033, through the grants PID2020-113334GB-I00 and PID2023-151418NB-I00. ASM also acknowledges the COIMBRA group scholarship for financial support.
 GE acknowledges support from the Austrian Science Fund FWF under grant number 10.55776/PAT2089624.
This work contributes to the aims of the USDOE ExoHad Topical Collaboration, contract DE-SC0023598. }

\providecommand{\href}[2]{#2}\begingroup\raggedright
\endgroup


\begin{thebibliography}{10}
  \setlength{\itemsep}{0mm}

\bibitem{Itzykson:1980rh}
C.~Itzykson and J.~B. Zuber, {\em {Quantum Field Theory}}.
\newblock International Series In Pure and Applied Physics. McGraw-Hill New
  York 1980

\bibitem{PrimEx-II:2020jwd}
{\bf PrimEx-II} Collaboration: I.~Larin {\em et al.}, ``\textit{{Precision
  measurement of the neutral pion lifetime}},''
  \href{http://dx.doi.org/10.1126/science.aay6641}{{\em Science} {\bf 368}
  (2020) no.~6490, 506--509}

\bibitem{Friedrich:2023kgv}
{\bf COMPASS} Collaboration: J.~Friedrich, ``\textit{{Chiral symmetry breaking:
  Current experimental status and prospects}},''
  \href{http://dx.doi.org/10.1051/epjconf/202328201007}{{\em EPJ Web Conf.}
  {\bf 282} (2023)  01007}

\bibitem{Ecker:2023qae}
{\bf COMPASS} Collaboration: D.~Ecker, ``\textit{{Testing predictions of the chiral anomaly in Primakoff reactions at COMPASS}},''
\href{https://doi.org/10.1393/ncc/i2024-24217-6}{{\em Nuovo Cim. C} {\bf 47} (2024) no.4, 217}


\bibitem{JFprivate}
J.~Friedrich and D.~Ecker, {\em {private discussion}}

\bibitem{Antipov:1986tp}
Y.~M. Antipov {\em et al.}, ``\textit{{Investigation of $\gamma \to 3 \pi$
  Chiral Anomaly During Pion Pair Production by Pions in the Nuclear Coulomb
  Field}},'' \href{http://dx.doi.org/10.1103/PhysRevD.36.21}{{\em Phys. Rev. D}
  {\bf 36} (1987)  21}

\bibitem{Moinester:2024lwl}
M.~Moinester, ``\textit{{Tribute to Henry Primakoff: Tests of Chiral
  Perturbation Theory via Primakoff Reactions}},''
  \href{http://arxiv.org/abs/2412.03669}{{\tt 2412.03669[hep-ph]}}
  \href{http://arxiv.org/abs/2412.03669}{{ $\bullet$}}
  \href{http://inspirehep.net/search?p=find+eprint+2412.03669}{{$
  \triangleright $}}

\bibitem{Niehus:2021iin}
M.~Niehus, M.~Hoferichter, and B.~Kubis, ``\textit{{The $\gamma\pi\to\pi\pi$
  anomaly from lattice QCD and dispersion relations}},''
  \href{http://dx.doi.org/10.1007/JHEP12(2021)038}{{\em JHEP} {\bf 12} (2021)
  038} \href{http://arxiv.org/abs/2110.11372}{{ $\bullet$}}
  \href{http://inspirehep.net/search?p=find+eprint+2110.11372}{{$
  \triangleright $}}

\bibitem{Hoferichter:2025lcz}
M.~Hoferichter, B.-L. Hoid, and B.~Kubis, ``\textit{{Extracting the chiral
  anomaly from $e^+e^-\to 3\pi$}},''
  \href{http://arxiv.org/abs/2504.13827}{{\tt 2504.13827[hep-ph]}}
  \href{http://arxiv.org/abs/2504.13827}{{ $\bullet$}}
  \href{http://inspirehep.net/search?p=find+eprint+2504.13827}{{$
  \triangleright $}}

\bibitem{Aoyama:2020ynm}
T.~Aoyama {\em et al.}, ``\textit{{The anomalous magnetic moment of the muon in
  the Standard Model}},''
  \href{http://dx.doi.org/10.1016/j.physrep.2020.07.006}{{\em Phys. Rept.} {\bf
  887} (2020)  1--166} \href{http://arxiv.org/abs/2006.04822}{{ $\bullet$}}
  \href{http://inspirehep.net/search?p=find+eprint+2006.04822}{{$
  \triangleright $}}

\bibitem{Colangelo:2022jxc}
G.~Colangelo {\em et al.}, ``\textit{{Prospects for precise predictions of
  $a_\mu$ in the Standard Model}},''
  \href{http://arxiv.org/abs/2203.15810}{{\tt 2203.15810[hep-ph]}}
  \href{http://arxiv.org/abs/2203.15810}{{ $\bullet$}}
  \href{http://inspirehep.net/search?p=find+eprint+2203.15810}{{$
  \triangleright $}}

\bibitem{Jain:1993qh}
P.~Jain and H.~J. Munczek, ``\textit{{q anti-q bound states in the
  Bethe-Salpeter formalism}},''
  \href{http://dx.doi.org/10.1103/PhysRevD.48.5403}{{\em Phys. Rev. D} {\bf 48}
  (1993)  5403--5411} \href{http://arxiv.org/abs/hep-ph/9307221}{{ $\bullet$}}
  \href{http://inspirehep.net/search?p=find+eprint+hep-ph/9307221}{{$
  \triangleright $}}

\bibitem{Maris:1997tm}
P.~Maris and C.~D. Roberts, ``\textit{{Pi- and K meson Bethe-Salpeter
  amplitudes}},'' \href{http://dx.doi.org/10.1103/PhysRevC.56.3369}{{\em Phys.
  Rev. C} {\bf 56} (1997)  3369--3383}
  \href{http://arxiv.org/abs/nucl-th/9708029}{{ $\bullet$}}
  \href{http://inspirehep.net/search?p=find+eprint+nucl-th/9708029}{{$
  \triangleright $}}

\bibitem{Maris:1997hd}
P.~Maris, C.~D. Roberts, and P.~C. Tandy, ``\textit{{Pion mass and decay
  constant}},'' \href{http://dx.doi.org/10.1016/S0370-2693(97)01535-9}{{\em
  Phys. Lett. B} {\bf 420} (1998)  267--273}
  \href{http://arxiv.org/abs/nucl-th/9707003}{{ $\bullet$}}
  \href{http://inspirehep.net/search?p=find+eprint+nucl-th/9707003}{{$
  \triangleright $}}

\bibitem{Maris:2000sk}
P.~Maris and P.~C. Tandy, ``\textit{{The pi, K+, and K0 electromagnetic
  form-factors}},'' \href{http://dx.doi.org/10.1103/PhysRevC.62.055204}{{\em
  Phys. Rev. C} {\bf 62} (2000)  055204}
  \href{http://arxiv.org/abs/nucl-th/0005015}{{ $\bullet$}}
  \href{http://inspirehep.net/search?p=find+eprint+nucl-th/0005015}{{$
  \triangleright $}}

\bibitem{Alkofer:2000wg}
R.~Alkofer and L.~von Smekal, ``\textit{{The Infrared behavior of QCD Green's
  functions: Confinement dynamical symmetry breaking, and hadrons as
  relativistic bound states}},''
  \href{http://dx.doi.org/10.1016/S0370-1573(01)00010-2}{{\em Phys. Rept.} {\bf
  353} (2001)  281} \href{http://arxiv.org/abs/hep-ph/0007355}{{ $\bullet$}}
  \href{http://inspirehep.net/search?p=find+eprint+hep-ph/0007355}{{$
  \triangleright $}}

\bibitem{Alkofer:2002bp}
R.~Alkofer, P.~Watson, and H.~Weigel, ``\textit{{Mesons in a Poincare covariant
  Bethe-Salpeter approach}},''
  \href{http://dx.doi.org/10.1103/PhysRevD.65.094026}{{\em Phys. Rev. D} {\bf
  65} (2002)  094026} \href{http://arxiv.org/abs/hep-ph/0202053}{{ $\bullet$}}
  \href{http://inspirehep.net/search?p=find+eprint+hep-ph/0202053}{{$
  \triangleright $}}

\bibitem{Fischer:2008wy}
C.~S. Fischer and R.~Williams, ``\textit{{Beyond the rainbow: Effects from pion
  back-coupling}},'' \href{http://dx.doi.org/10.1103/PhysRevD.78.074006}{{\em
  Phys. Rev. D} {\bf 78} (2008)  074006}
  \href{http://arxiv.org/abs/0808.3372}{{ $\bullet$}}
  \href{http://inspirehep.net/search?p=find+eprint+0808.3372}{{$ \triangleright
  $}}

\bibitem{Krassnigg:2008bob}
A.~Krassnigg, ``\textit{{Excited mesons in a Bethe-Salpeter approach}},''
  \href{http://dx.doi.org/10.22323/1.077.0075}{{\em PoS} {\bf CONFINEMENT8}
  (2008)  075} \href{http://arxiv.org/abs/0812.3073}{{ $\bullet$}}
  \href{http://inspirehep.net/search?p=find+eprint+0812.3073}{{$ \triangleright
  $}}

\bibitem{Qin:2011dd}
S.-x. Qin, L.~Chang, Y.-x. Liu, C.~D. Roberts, and D.~J. Wilson,
  ``\textit{{Interaction model for the gap equation}},''
  \href{http://dx.doi.org/10.1103/PhysRevC.84.042202}{{\em Phys. Rev. C} {\bf
  84} (2011)  042202} \href{http://arxiv.org/abs/1108.0603}{{ $\bullet$}}
  \href{http://inspirehep.net/search?p=find+eprint+1108.0603}{{$ \triangleright
  $}}

\bibitem{Binosi:2014aea}
D.~Binosi, L.~Chang, J.~Papavassiliou, and C.~D. Roberts, ``\textit{{Bridging a
  gap between continuum-QCD and ab initio predictions of hadron
  observables}},'' \href{http://dx.doi.org/10.1016/j.physletb.2015.01.031}{{\em
  Phys. Lett. B} {\bf 742} (2015)  183--188}
  \href{http://arxiv.org/abs/1412.4782}{{ $\bullet$}}
  \href{http://inspirehep.net/search?p=find+eprint+1412.4782}{{$ \triangleright
  $}}

\bibitem{Rojas:2014aka}
E.~Rojas, B.~El-Bennich, and J.~P. B.~C. de~Melo, ``\textit{{Exciting flavored
  bound states}},'' \href{http://dx.doi.org/10.1103/PhysRevD.90.074025}{{\em
  Phys. Rev. D} {\bf 90} (2014)  074025}
  \href{http://arxiv.org/abs/1407.3598}{{ $\bullet$}}
  \href{http://inspirehep.net/search?p=find+eprint+1407.3598}{{$ \triangleright
  $}}

\bibitem{Binosi:2016rxz}
D.~Binosi, L.~Chang, J.~Papavassiliou, S.-X. Qin, and C.~D. Roberts,
  ``\textit{{Symmetry preserving truncations of the gap and Bethe-Salpeter
  equations}},'' \href{http://dx.doi.org/10.1103/PhysRevD.93.096010}{{\em Phys.
  Rev. D} {\bf 93} (2016) no.~9, 096010}
  \href{http://arxiv.org/abs/1601.05441}{{ $\bullet$}}
  \href{http://inspirehep.net/search?p=find+eprint+1601.05441}{{$
  \triangleright $}}

\bibitem{El-Bennich:2016qmb}
B.~El-Bennich, G.~a. Krein, E.~Rojas, and F.~E. Serna, ``\textit{{Excited
  hadrons and the analytical structure of bound-state interaction kernels}},''
  \href{http://dx.doi.org/10.1007/s00601-016-1133-x}{{\em Few Body Syst.} {\bf
  57} (2016) no.~10, 955--963} \href{http://arxiv.org/abs/1602.06761}{{
  $\bullet$}} \href{http://inspirehep.net/search?p=find+eprint+1602.06761}{{$
  \triangleright $}}

\bibitem{Eichmann:2016yit}
G.~Eichmann, H.~Sanchis-Alepuz, R.~Williams, R.~Alkofer, and C.~S. Fischer,
  ``\textit{{Baryons as relativistic three-quark bound states}},''
  \href{http://dx.doi.org/10.1016/j.ppnp.2016.07.001}{{\em Prog. Part. Nucl.
  Phys.} {\bf 91} (2016)  1--100} \href{http://arxiv.org/abs/1606.09602}{{
  $\bullet$}} \href{http://inspirehep.net/search?p=find+eprint+1606.09602}{{$
  \triangleright $}}

\bibitem{Xu:2022kng}
Z.-N. Xu, Z.-Q. Yao, S.-X. Qin, Z.-F. Cui, and C.~D. Roberts,
  ``\textit{{Bethe\textendash{}Salpeter kernel and properties of strange-quark
  mesons}},'' \href{http://dx.doi.org/10.1140/epja/s10050-023-00951-7}{{\em
  Eur. Phys. J. A} {\bf 59} (2023) no.~3, 39}
  \href{http://arxiv.org/abs/2208.13903}{{ $\bullet$}}
  \href{http://inspirehep.net/search?p=find+eprint+2208.13903}{{$
  \triangleright $}}

\bibitem{Miramontes:2019mco}
A.~S. Miramontes and H.~Sanchis-Alepuz, ``\textit{{On the effect of resonances
  in the quark-photon vertex}},''
  \href{http://dx.doi.org/10.1140/epja/i2019-12847-6}{{\em Eur. Phys. J. A}
  {\bf 55} (2019) no.~10, 170} \href{http://arxiv.org/abs/1906.06227}{{
  $\bullet$}} \href{http://inspirehep.net/search?p=find+eprint+1906.06227}{{$
  \triangleright $}}

\bibitem{Eichmann:2020oqt}
G.~Eichmann, C.~S. Fischer, W.~Heupel, N.~Santowsky, and P.~C. Wallbott,
  ``\textit{{Four-Quark States from Functional Methods}},''
  \href{http://dx.doi.org/10.1007/s00601-020-01571-3}{{\em Few Body Syst.} {\bf
  61} (2020) no.~4, 38} \href{http://arxiv.org/abs/2008.10240}{{ $\bullet$}}
  \href{http://inspirehep.net/search?p=find+eprint+2008.10240}{{$
  \triangleright $}}

\bibitem{Miramontes:2021xgn}
A.~S. Miramontes, H.~Sanchis~Alepuz, and R.~Alkofer, ``\textit{{Elucidating the
  effect of intermediate resonances in the quark interaction kernel on the
  timelike electromagnetic pion form factor}},''
  \href{http://dx.doi.org/10.1103/PhysRevD.103.116006}{{\em Phys. Rev. D} {\bf
  103} (2021) no.~11, 116006} \href{http://arxiv.org/abs/2102.12541}{{
  $\bullet$}} \href{http://inspirehep.net/search?p=find+eprint+2102.12541}{{$
  \triangleright $}}

\bibitem{Huber:2020ngt}
M.~Q. Huber, C.~S. Fischer, and H.~Sanchis-Alepuz, ``\textit{{Spectrum of
  scalar and pseudoscalar glueballs from functional methods}},''
  \href{http://dx.doi.org/10.1140/epjc/s10052-020-08649-6}{{\em Eur. Phys. J.
  C} {\bf 80} (2020) no.~11, 1077} \href{http://arxiv.org/abs/2004.00415}{{
  $\bullet$}} \href{http://inspirehep.net/search?p=find+eprint+2004.00415}{{$
  \triangleright $}}

\bibitem{Huber:2021yfy}
M.~Q. Huber, C.~S. Fischer, and H.~Sanchis-Alepuz, ``\textit{{Higher spin
  glueballs from functional methods}},''
  \href{http://dx.doi.org/10.1140/epjc/s10052-021-09864-5}{{\em Eur. Phys. J.
  C} {\bf 81} (2021) no.~12, 1083} \href{http://arxiv.org/abs/2110.09180}{{
  $\bullet$}} \href{http://inspirehep.net/search?p=find+eprint+2110.09180}{{$
  \triangleright $}}

\bibitem{Eichmann:2023tjk}
G.~Eichmann, A.~G\'omez, J.~Horak, J.~M. Pawlowski, J.~Wessely, and N.~Wink,
  ``\textit{{Bound states from the spectral Bethe-Salpeter equation}},''
  \href{http://dx.doi.org/10.1103/PhysRevD.109.096024}{{\em Phys. Rev. D} {\bf
  109} (2024) no.~9, 096024} \href{http://arxiv.org/abs/2310.16353}{{
  $\bullet$}} \href{http://inspirehep.net/search?p=find+eprint+2310.16353}{{$
  \triangleright $}}

\bibitem{Gao:2024gdj}
F.~Gao, A.~S. Miramontes, J.~Papavassiliou, and J.~M. Pawlowski,
  ``\textit{{Heavy-light mesons from a flavour-dependent interaction}},''
  \href{http://dx.doi.org/10.1016/j.physletb.2025.139384}{{\em Phys. Lett. B}
  {\bf 863} (2025)  139384} \href{http://arxiv.org/abs/2411.19680}{{
  $\bullet$}} \href{http://inspirehep.net/search?p=find+eprint+2411.19680}{{$
  \triangleright $}}

\bibitem{Eichmann:2017wil}
G.~Eichmann, C.~S. Fischer, E.~Weil, and R.~Williams, ``\textit{{On the
  large-$Q^2$ behavior of the pion transition form factor}},''
  \href{http://dx.doi.org/10.1016/j.physletb.2017.10.002}{{\em Phys. Lett. B}
  {\bf 774} (2017)  425--429} \href{http://arxiv.org/abs/1704.05774}{{
  $\bullet$}} \href{http://inspirehep.net/search?p=find+eprint+1704.05774}{{$
  \triangleright $}}

\bibitem{Weil:2017knt}
E.~Weil, G.~Eichmann, C.~S. Fischer, and R.~Williams,
  ``\textit{{Electromagnetic decays of the neutral pion}},''
  \href{http://dx.doi.org/10.1103/PhysRevD.96.014021}{{\em Phys. Rev. D} {\bf
  96} (2017) no.~1, 014021} \href{http://arxiv.org/abs/1704.06046}{{
  $\bullet$}} \href{http://inspirehep.net/search?p=find+eprint+1704.06046}{{$
  \triangleright $}}

\bibitem{Raya:2019dnh}
K.~Raya, A.~Bashir, and P.~Roig, ``\textit{{Contribution of neutral
  pseudoscalar mesons to $a_\mu^{HLbL}$ within a Schwinger-Dyson equations
  approach to QCD}},''
  \href{http://dx.doi.org/10.1103/PhysRevD.101.074021}{{\em Phys. Rev. D} {\bf
  101} (2020) no.~7, 074021} \href{http://arxiv.org/abs/1910.05960}{{
  $\bullet$}} \href{http://inspirehep.net/search?p=find+eprint+1910.05960}{{$
  \triangleright $}}

\bibitem{Eichmann:2019tjk}
G.~Eichmann, C.~S. Fischer, E.~Weil, and R.~Williams, ``\textit{{Single
  pseudoscalar meson pole and pion box contributions to the anomalous magnetic
  moment of the muon}},''
  \href{http://dx.doi.org/10.1016/j.physletb.2019.134855}{{\em Phys. Lett. B}
  {\bf 797} (2019)  134855} \href{http://arxiv.org/abs/1903.10844}{{
  $\bullet$}} \href{http://inspirehep.net/search?p=find+eprint+1903.10844}{{$
  \triangleright $}}

\bibitem{Eichmann:2024glq}
G.~Eichmann, C.~S. Fischer, T.~Haeuser, and O.~Regenfelder,
  ``\textit{{Axial-vector and scalar contributions to hadronic light-by-light
  scattering}},'' \href{http://arxiv.org/abs/2411.05652}{{\tt
  2411.05652[hep-ph]}} \href{http://arxiv.org/abs/2411.05652}{{ $\bullet$}}
  \href{http://inspirehep.net/search?p=find+eprint+2411.05652}{{$
  \triangleright $}}

\bibitem{Alkofer:1995jx}
R.~Alkofer and C.~D. Roberts, ``\textit{{Calculation of the anomalous gamma pi*
  --\ensuremath{>} pi pi form-factor}},''
  \href{http://dx.doi.org/10.1016/0370-2693(95)01517-5}{{\em Phys. Lett. B}
  {\bf 369} (1996)  101--107} \href{http://arxiv.org/abs/hep-ph/9510284}{{
  $\bullet$}}
  \href{http://inspirehep.net/search?p=find+eprint+hep-ph/9510284}{{$
  \triangleright $}}

\bibitem{Cotanch:2003xv}
S.~R. Cotanch and P.~Maris, ``\textit{{Ladder Dyson-Schwinger calculation of
  the anomalous gamma-3pi form-factor}},''
  \href{http://dx.doi.org/10.1103/PhysRevD.68.036006}{{\em Phys. Rev. D} {\bf
  68} (2003)  036006} \href{http://arxiv.org/abs/nucl-th/0308008}{{ $\bullet$}}
  \href{http://inspirehep.net/search?p=find+eprint+nucl-th/0308008}{{$
  \triangleright $}}

\bibitem{Bistrovic:1999dy}
B.~Bistrovic and D.~Klabucar, ``\textit{{Anomalous gamma ---\ensuremath{>} 3 pi
  amplitude in a bound state approach}},''
  \href{http://dx.doi.org/10.1016/S0370-2693(00)00241-0}{{\em Phys. Lett. B}
  {\bf 478} (2000)  127--136} \href{http://arxiv.org/abs/hep-ph/9912452}{{
  $\bullet$}}
  \href{http://inspirehep.net/search?p=find+eprint+hep-ph/9912452}{{$
  \triangleright $}}

\bibitem{Xing:2024bpj}
Z.~Xing, H.~Dang, M.~A. Sultan, K.~Raya, and L.~Chang, ``\textit{{QCD anomalies
  in electromagnetic processes: A solution to the
  \ensuremath{\gamma}\textrightarrow{}3\ensuremath{\pi} puzzle}},''
  \href{http://dx.doi.org/10.1103/PhysRevD.109.054028}{{\em Phys. Rev. D} {\bf
  109} (2024) no.~5, 054028} \href{http://arxiv.org/abs/2401.03169}{{
  $\bullet$}} \href{http://inspirehep.net/search?p=find+eprint+2401.03169}{{$
  \triangleright $}}

\bibitem{Adler:1971nq}
S.~L. Adler, B.~W. Lee, S.~B. Treiman, and A.~Zee, ``\textit{{Low Energy
  Theorem for $\gamma^+ \gamma \to \pi^+ \pi^+ \pi$}},''
  \href{http://dx.doi.org/10.1103/PhysRevD.4.3497}{{\em Phys. Rev. D} {\bf 4}
  (1971)  3497--3501}

\bibitem{Aviv:1971hq}
R.~Aviv and A.~Zee, ``\textit{{Low-energy theorem for gamma --\ensuremath{>} 3
  pi}},'' \href{http://dx.doi.org/10.1103/PhysRevD.5.2372}{{\em Phys. Rev. D}
  {\bf 5} (1972)  2372}

\bibitem{Bashir:2012fs}
A.~Bashir, L.~Chang, I.~C. Cloet, B.~El-Bennich, Y.-X. Liu, C.~D. Roberts, and
  P.~C. Tandy, ``\textit{{Collective perspective on advances in Dyson-Schwinger
  Equation QCD}},'' \href{http://dx.doi.org/10.1088/0253-6102/58/1/16}{{\em
  Commun. Theor. Phys.} {\bf 58} (2012)  79--134}
  \href{http://arxiv.org/abs/1201.3366}{{ $\bullet$}}
  \href{http://inspirehep.net/search?p=find+eprint+1201.3366}{{$ \triangleright
  $}}

\bibitem{Sanchis-Alepuz:2017jjd}
H.~Sanchis-Alepuz and R.~Williams, ``\textit{{Recent developments in
  bound-state calculations using the Dyson\textendash{}Schwinger and
  Bethe\textendash{}Salpeter equations}},''
  \href{http://dx.doi.org/10.1016/j.cpc.2018.05.020}{{\em Comput. Phys.
  Commun.} {\bf 232} (2018)  1--21} \href{http://arxiv.org/abs/1710.04903}{{
  $\bullet$}} \href{http://inspirehep.net/search?p=find+eprint+1710.04903}{{$
  \triangleright $}}

\bibitem{Frank:1994mf}
M.~R. Frank, ``\textit{{Nonperturbative aspects of the quark - photon
  vertex}},'' \href{http://dx.doi.org/10.1103/PhysRevC.51.987}{{\em Phys. Rev.
  C} {\bf 51} (1995)  987--998} \href{http://arxiv.org/abs/nucl-th/9403009}{{
  $\bullet$}}
  \href{http://inspirehep.net/search?p=find+eprint+nucl-th/9403009}{{$
  \triangleright $}}

\bibitem{Maris:1999bh}
P.~Maris and P.~C. Tandy, ``\textit{{The Quark photon vertex and the pion
  charge radius}},'' \href{http://dx.doi.org/10.1103/PhysRevC.61.045202}{{\em
  Phys. Rev. C} {\bf 61} (2000)  045202}
  \href{http://arxiv.org/abs/nucl-th/9910033}{{ $\bullet$}}
  \href{http://inspirehep.net/search?p=find+eprint+nucl-th/9910033}{{$
  \triangleright $}}

\bibitem{Eichmann:2014qva}
G.~Eichmann, ``\textit{{Probing nucleons with photons at the quark level}},''
  \href{http://dx.doi.org/10.5506/APhysPolBSupp.7.597}{{\em Acta Phys. Polon.
  Supp.} {\bf 7} (2014) no.~3, 597} \href{http://arxiv.org/abs/1404.4149}{{
  $\bullet$}} \href{http://inspirehep.net/search?p=find+eprint+1404.4149}{{$
  \triangleright $}}

\bibitem{Leutnant:2018dry}
M.~Leutnant and A.~Sternbeck, ``\textit{{Quark-photon vertex from lattice QCD
  in Landau gauge}},'' \href{http://dx.doi.org/10.22323/1.336.0095}{{\em PoS}
  {\bf Confinement2018} (2018)  095} \href{http://arxiv.org/abs/1812.11131}{{
  $\bullet$}} \href{http://inspirehep.net/search?p=find+eprint+1812.11131}{{$
  \triangleright $}}

\bibitem{Tang:2019zbk}
C.~Tang, F.~Gao, and Y.-X. Liu, ``\textit{{Practical scheme from QCD to
  phenomena via Dyson-Schwinger equations}},''
  \href{http://dx.doi.org/10.1103/PhysRevD.100.056001}{{\em Phys. Rev. D} {\bf
  100} (2019) no.~5, 056001} \href{http://arxiv.org/abs/1902.01679}{{
  $\bullet$}} \href{http://inspirehep.net/search?p=find+eprint+1902.01679}{{$
  \triangleright $}}

\bibitem{Ball:1980ay}
J.~S. Ball and T.-W. Chiu, ``\textit{{Analytic Properties of the Vertex
  Function in Gauge Theories. 1.}},''
  \href{http://dx.doi.org/10.1103/PhysRevD.22.2542}{{\em Phys. Rev. D} {\bf 22}
  (1980)  2542}

\bibitem{Fischer:2007ze}
C.~S. Fischer, D.~Nickel, and J.~Wambach, ``\textit{{Hadronic unquenching
  effects in the quark propagator}},''
  \href{http://dx.doi.org/10.1103/PhysRevD.76.094009}{{\em Phys. Rev. D} {\bf
  76} (2007)  094009} \href{http://arxiv.org/abs/0705.4407}{{ $\bullet$}}
  \href{http://inspirehep.net/search?p=find+eprint+0705.4407}{{$ \triangleright
  $}}

\bibitem{Fischer:2008sp}
C.~S. Fischer, D.~Nickel, and R.~Williams, ``\textit{{On Gribov's
  supercriticality picture of quark confinement}},''
  \href{http://dx.doi.org/10.1140/epjc/s10052-008-0821-1}{{\em Eur. Phys. J. C}
  {\bf 60} (2009)  47--61} \href{http://arxiv.org/abs/0807.3486}{{ $\bullet$}}
  \href{http://inspirehep.net/search?p=find+eprint+0807.3486}{{$ \triangleright
  $}}

\bibitem{Williams:2018adr}
R.~Williams, ``\textit{{Vector mesons as dynamical resonances in the
  Bethe\textendash{}Salpeter framework}},''
  \href{http://dx.doi.org/10.1016/j.physletb.2019.134943}{{\em Phys. Lett. B}
  {\bf 798} (2019)  134943} \href{http://arxiv.org/abs/1804.11161}{{
  $\bullet$}} \href{http://inspirehep.net/search?p=find+eprint+1804.11161}{{$
  \triangleright $}}

\bibitem{Alkofer:2022hln}
R.~Alkofer, A.~S. Miramontes, and H.~Sanchis-Alepuz, ``\textit{{Elucidating the
  \ensuremath{\rho}-meson\textquoteright{}s role as intermediate resonance in
  the time-like electromagnetic pion form factor}},''
  \href{http://dx.doi.org/10.1051/epjconf/202226201020}{{\em EPJ Web Conf.}
  {\bf 262} (2022)  01020} \href{http://arxiv.org/abs/2202.05056}{{ $\bullet$}}
  \href{http://inspirehep.net/search?p=find+eprint+2202.05056}{{$
  \triangleright $}}

\bibitem{Cotanch:2002vj}
S.~R. Cotanch and P.~Maris, ``\textit{{QCD based quark description of pi pi
  scattering up to the sigma and rho region}},''
  \href{http://dx.doi.org/10.1103/PhysRevD.66.116010}{{\em Phys. Rev. D} {\bf
  66} (2002)  116010} \href{http://arxiv.org/abs/hep-ph/0210151}{{ $\bullet$}}
  \href{http://inspirehep.net/search?p=find+eprint+hep-ph/0210151}{{$
  \triangleright $}}

\bibitem{Goecke:2012qm}
T.~Goecke, C.~S. Fischer, and R.~Williams, ``\textit{{Role of momentum
  dependent dressing functions and vector meson dominance in hadronic
  light-by-light contributions to the muon $g-2$}},''
  \href{http://dx.doi.org/10.1103/PhysRevD.87.034013}{{\em Phys. Rev. D} {\bf
  87} (2013) no.~3, 034013} \href{http://arxiv.org/abs/1210.1759}{{ $\bullet$}}
  \href{http://inspirehep.net/search?p=find+eprint+1210.1759}{{$ \triangleright
  $}}

\bibitem{Eichmann:2011ec}
G.~Eichmann and C.~S. Fischer, ``\textit{{Unified description of hadron-photon
  and hadron-meson scattering in the Dyson-Schwinger approach}},''
  \href{http://dx.doi.org/10.1103/PhysRevD.85.034015}{{\em Phys. Rev. D} {\bf
  85} (2012)  034015} \href{http://arxiv.org/abs/1111.0197}{{ $\bullet$}}
  \href{http://inspirehep.net/search?p=find+eprint+1111.0197}{{$ \triangleright
  $}}

\bibitem{Eichmann:2012mp}
G.~Eichmann and C.~S. Fischer, ``\textit{{Nucleon Compton scattering in the
  Dyson-Schwinger approach}},''
  \href{http://dx.doi.org/10.1103/PhysRevD.87.036006}{{\em Phys. Rev. D} {\bf
  87} (2013) no.~3, 036006} \href{http://arxiv.org/abs/1212.1761}{{ $\bullet$}}
  \href{http://inspirehep.net/search?p=find+eprint+1212.1761}{{$ \triangleright
  $}}

\bibitem{Eichmann:2014ooa}
G.~Eichmann, C.~S. Fischer, W.~Heupel, and R.~Williams, ``\textit{{The muon
  g-2: Dyson-Schwinger status on hadronic light-by-light scattering}},''
  \href{http://dx.doi.org/10.1063/1.4938621}{{\em AIP Conf. Proc.} {\bf 1701}
  (2016) no.~1, 040004} \href{http://arxiv.org/abs/1411.7876}{{ $\bullet$}}
  \href{http://inspirehep.net/search?p=find+eprint+1411.7876}{{$ \triangleright
  $}}

\bibitem{AnDiPrep}
D.~An, G.~Eichmann, C.~S. Fischer, and S.~Leupold, {\em in preparation}

\bibitem{ParticleDataGroup:2024cfk}
{\bf Particle Data Group} Collaboration: S.~Navas {\em et al.},
  ``\textit{{Review of particle physics}},''
  \href{http://dx.doi.org/10.1103/PhysRevD.110.030001}{{\em Phys. Rev. D} {\bf
  110} (2024) no.~3, 030001}

\bibitem{Eichmann:2014xya}
G.~Eichmann, R.~Williams, R.~Alkofer, and M.~Vujinovic, ``\textit{Three-gluon
  vertex in landau gauge},''
  \href{http://dx.doi.org/10.1103/PhysRevD.89.105014}{{\em Phys. Rev. D} {\bf
  89} (2014) no.~10, 105014} \href{http://arxiv.org/abs/1402.1365}{{
  $\bullet$}} \href{http://inspirehep.net/search?p=find+eprint+1402.1365}{{$
  \triangleright $}}

\bibitem{Eichmann:2015nra}
G.~Eichmann, C.~S. Fischer, and W.~Heupel, ``\textit{Four-point functions and
  the permutation group s4},''
  \href{http://dx.doi.org/10.1103/PhysRevD.92.056006}{{\em Phys. Rev. D} {\bf
  92} (2015) no.~5, 056006} \href{http://arxiv.org/abs/1505.06336}{{
  $\bullet$}} \href{http://inspirehep.net/search?p=find+eprint+1505.06336}{{$
  \triangleright $}}

\bibitem{Pinto-Gomez:2022brg}
F.~Pinto-G\'omez, F.~De~Soto, M.~N. Ferreira, J.~Papavassiliou, and
  J.~Rodr\'\i{}guez-Quintero, ``\textit{Lattice three-gluon vertex in extended
  kinematics: Planar degeneracy},''
  \href{http://dx.doi.org/10.1016/j.physletb.2023.137737}{{\em Phys. Lett. B}
  {\bf 838} (2023)  137737} \href{http://arxiv.org/abs/2208.01020}{{
  $\bullet$}} \href{http://inspirehep.net/search?p=find+eprint+2208.01020}{{$
  \triangleright $}}

\bibitem{Ferreira:2023fva}
M.~N. Ferreira and J.~Papavassiliou, ``\textit{Gauge sector dynamics in qcd},''
  \href{http://dx.doi.org/10.3390/particles6010017}{{\em Particles} {\bf 6}
  (2023) no.~1, 312--363} \href{http://arxiv.org/abs/2301.02314}{{ $\bullet$}}
  \href{http://inspirehep.net/search?p=find+eprint+2301.02314}{{$
  \triangleright $}}

\bibitem{Aguilar:2023qqd}
A.~C. Aguilar, M.~N. Ferreira, J.~Papavassiliou, and L.~R. Santos,
  ``\textit{Planar degeneracy of the three-gluon vertex},''
  \href{http://dx.doi.org/10.1140/epjc/s10052-023-11732-3}{{\em Eur. Phys. J.
  C} {\bf 83} (2023) no.~6, 549} \href{http://arxiv.org/abs/2305.05704}{{
  $\bullet$}} \href{http://inspirehep.net/search?p=find+eprint+2305.05704}{{$
  \triangleright $}}

\bibitem{Aguilar:2024fen}
A.~C. Aguilar, M.~N. Ferreira, J.~Papavassiliou, and L.~R. Santos,
  ``\textit{Four-gluon vertex in collinear kinematics},''
  \href{http://dx.doi.org/10.1140/epjc/s10052-024-12970-9}{{\em Eur. Phys. J.
  C} {\bf 84} (2024) no.~7, 676} \href{http://arxiv.org/abs/2402.16071}{{
  $\bullet$}} \href{http://inspirehep.net/search?p=find+eprint+2402.16071}{{$
  \triangleright $}}

\bibitem{Briceno:2016kkp}
R.~A. Brice\~no, J.~J. Dudek, R.~G. Edwards, C.~J. Shultz, C.~E. Thomas, and
  D.~J. Wilson, ``\textit{{The $\pi\pi\to\pi\gamma^\star$ amplitude and the
  resonant $\rho\to\pi\gamma^\star$ transition from lattice QCD}},''
  \href{http://dx.doi.org/10.1103/PhysRevD.93.114508}{{\em Phys. Rev. D} {\bf
  93} (2016) no.~11, 114508} \href{http://arxiv.org/abs/1604.03530}{{
  $\bullet$}} \href{http://inspirehep.net/search?p=find+eprint+1604.03530}{{$
  \triangleright $}}

\bibitem{Hoferichter:2012pm}
M.~Hoferichter, B.~Kubis, and D.~Sakkas, ``\textit{{Extracting the chiral
  anomaly from gamma pi --\ensuremath{>} pi pi}},''
  \href{http://dx.doi.org/10.1103/PhysRevD.86.116009}{{\em Phys. Rev. D} {\bf
  86} (2012)  116009} \href{http://arxiv.org/abs/1210.6793}{{ $\bullet$}}
  \href{http://inspirehep.net/search?p=find+eprint+1210.6793}{{$ \triangleright
  $}}


\bibitem{Hoferichter:2017ftn}
M.~Hoferichter, B.~Kubis and M.~Zanke, ``\textit{{Radiative resonance couplings in \ensuremath{\gamma}\ensuremath{\pi}\textrightarrow{}\ensuremath{\pi}\ensuremath{\pi},}}''
\href{http://dx.doi.org/10.1103/PhysRevD.96.114016}
{\em Phys. Rev. D \bf{96} (2017) 114016}
\href{http://arxiv.org/abs/1210.6793}{{ $\bullet$}}
  \href{http://inspirehep.net/search?p=find+eprint+1710.00824}{{$ \triangleright
  $}}

\bibitem{Ecker:1994gg}
G.~Ecker, ``\textit{{Chiral perturbation theory}},''
  \href{http://dx.doi.org/10.1016/0146-6410(95)00041-G}{{\em Prog. Part. Nucl.
  Phys.} {\bf 35} (1995)  1--80} \href{http://arxiv.org/abs/hep-ph/9501357}{{
  $\bullet$}}
  \href{http://inspirehep.net/search?p=find+eprint+hep-ph/9501357}{{$
  \triangleright $}}

\bibitem{Bijnens:1989ff}
J.~Bijnens, A.~Bramon and F.~Cornet, ``\textit{{Three Pseudoscalar Photon Interactions in Chiral Perturbation Theory}}'',
\href{http://dx.doi.org/10.1016/0370-2693(90)91212-T}
{\em Phys. Lett. B \bf{237} (1990) 488}

\end{thebibliography}
\end{document}